\documentclass[11pt,preprint]{aastex}

\usepackage[dvips]{epsfig}
\usepackage{natbib}
\usepackage{emulateapj5}

\shorttitle{Continuous Energy Injection in GRB~010222}
\shortauthors{Bj\"ornsson et~al.}

\received{2002 September 2}
\begin{document}

\title{The Afterglow of GRB~010222: A Case of Continuous
Energy Injection\altaffilmark{1}}

\author{G.~Bj\"ornsson\altaffilmark{2}, J.~Hjorth\altaffilmark{3}, 
K.~Pedersen\altaffilmark{3} and J.U.~Fynbo\altaffilmark{4}}

\altaffiltext{1}{Based on observations made with the Nordic Optical Telescope,
operated on the island of La Palma jointly by Denmark, Finland, Iceland, Norway, 
and Sweden, in the Spanish Observatorio del Roque de los Muchachos of the 
Instituto de Astrofisica de Canarias.}
\altaffiltext{2}{Science Institute, University of Iceland, Dunhaga~3,
  IS--107 Reykjavik, Iceland, e-mail: gulli@raunvis.hi.is}
\altaffiltext{3}{Astronomical Observatory, University of Copenhagen,
  Juliane Maries Vej 30, DK--2100 Copenhagen \O, Denmark, e-mail: jens@astro.ku.dk,
  kp@astro.ku.dk}
\altaffiltext{4}{Department of Physics and Astronomy, Ny Munkegade,
  DK-8000 Aarhus C, Denmark, e-mail: jfynbo@phys.au.dk}

\begin{abstract}
The optical lightcurve of GRB~010222 exhibited one of the slowest decays of 
any gamma-ray burst to date. Its broadband properties have been difficult 
to explain with conventional afterglow models, as they either require 
the power law index of the underlying electron energy distribution 
to be low, $p<2$, or that the outflow is quasi-spherical thus reviving
the energy problem. We argue that the slow decay of GRB~010222 and a linear 
polarization of $1.36\pm 0.64$\%, is naturally explained by a jet model with 
continuous energy injection. The electron energy distribution then has 
$p=2.49\pm0.05$, fully consistent with the expectation from detailed 
modelling of acceleration in relativistic shocks, that $p>2$, thus 
alleviating the ``$p$-problem''.

\end{abstract}


\section{Introduction}
\label{sec:intro}

GRB~010222 was a bright gamma-ray burst (GRB) localized by BeppoSAX\\
\citep{piro01}. X-ray observations were reported 
by in 't Zand et al.\ (2001).  The optical afterglow 
was discovered by Henden \& Vrba \cite{henden01}, 4.3 hours after the 
burst and a redshift of $z=1.477$ was determined by Jha et al.\ \cite{jha01}.
Further optical/near-infrared observations have been reported by 
Stanek et al.\ \cite{stanek01}, Lee et al.\ \cite{lee01}, 
Masetti et al.\ \cite{masetti01}, Cowsik et al.\ (2001), 
Sagar et al.\ \cite{sagar01} and Mirabal et al.\ (2002). 
In our analysis we shall adopt 
the light curve fit of Henden et al.\ (2002), that is based 
on data extending up to 80 days after the burst. Their best fit in the 
$R$-band gives a pre-break power-law slope of $\alpha_1=-0.66\pm 0.03$ 
and a post-break slope of $\alpha_2=-1.40\pm 0.02$, with a break time 
of $t_b=0.58\pm 0.04$ days. The observed optical spectral index is 
about $\beta=-1.0$ (e.g.\ Stanek et al.\ 2001), but as it may be 
strongly affected by host extinction, it can lead to ambiguous 
inference of the afterglow properties. 
In particular, the relationship between the light curve decay
indices, the $\alpha$'s, and the spectral index, $\beta$, 
that is predicted by synchrotron models of afterglows (e.g.\
Sari, Piran \& Narayan 1998; Sari, Piran \& Halpern 1999; Price et al.\
2002), can lead to inconsistent interpretations due to the unknown host 
extinction.

To date, interpretation of the GRB~010222 afterglow observations has mainly 
relied on two possible scenarios: i) A narrow sideways expanding 
jet propagating in a low-density medium (e.g.\ Stanek et al.\ 2001). 
ii) A wide jet or spherical fireball transiting from relativistic to 
non-relativistic regime (e.g.\ Masetti et al.\ 2001; in 't Zand et al.\ 2001).
The latter models require a very dense medium and implies a very 
large energy release, thus bringing back the energy problem 
\citep{kulkarni99, andersen99}. Both classes of models require 
a hard electron energy distribution, i.e.\ $p<2$, in the first case 
even as low as $p\approx 1.4$. This is contrary to most other bursts 
that seem to be adequately fit with models where $p\approx 2.3-2.5$
(van Paradijs, Kouvelioutu \& Wijers 2000). 
Recently, Panaitescu \& Kumar \cite{pankum02} presented fits to several 
afterglows, concluding that $p$ is smaller than 2 in a number of them. 
Such small inferred values of $p$ signal a 
departure from the standard fireball model \citep{mesz02}, and introduce
additional free parameters into the model, such as an upper cutoff in the 
electron energy distribution, or additional assumptions about the electron 
acceleration mechanism to facilitate generation of flat energy 
distributions (Dai \& Cheng 2001; Bhattacharya 2001). The $p$-problem 
then arises from the fact that detailed modelling of particle acceleration 
in relativistic shocks indicates that $p\approx 2.2-2.3$ \citep{achterberg01}.

In addition, if $p$ is assumed to be constant throughout 
the fireball evolution, the magnitude of the observed light curve break
in the case of GRB~010222, $\Delta\alpha=\alpha_1-\alpha_2=0.74\pm 0.04$, 
cannot be explained by the above models. The sideways expanding jet model 
predicts a break magnitude of $\Delta\alpha=1-\alpha_1/3=1.22\pm 0.01$, 
while the fireball transiting to the non-relativistic regime gives 
$\Delta\alpha=(\alpha_1+3/5)=0.06\pm 0.03$ for slow cooling 
electrons. For fast cooling electrons the latter model predicts 
$\Delta\alpha=(\alpha_1+1)=0.34\pm 0.03$. 
The break magnitudes predicted by the models are in all cases very 
different from the observed value that is however, in perfect agreement 
with the prediction of a jet model with a fixed opening angle \citep{mesz99}, 
$\Delta\alpha=3/4$, being essentially of geometrical origin.

In this Letter we use polarization measurements to argue against 
interpreatations based on spherical models transiting to the non-relativistic 
regime. Furthermore, we demonstrate that the observations of GRB~010222 
can be naturally interpreted with a jet model with a small opening angle 
and continuous energy injection. 

\section{Observations}
\label{sec:obs}

\subsection{Polarimetry}
Polarimetric imaging observations were obtained at the 2.56-m Nordic
Optical Telescope (NOT) on La Palma, The Canary Islands, with the
Andaluc\'{\i}a faint object spectrograph (ALFOSC), using two calcite
plates together with a $V$-band filter. Each calcite plate provides the
simultaneous measurement of the ordinary and the extraordinary components
of two orthogonally polarized beams. Thus, one image gives either the
0$^\circ$ and 90$^\circ$ components or the 45$^\circ$ and 135$^\circ$
components simultaneously.  The calcite plates produce a vignetted field
of about 140\arcsec\ in diameter. The detector was a thinned Loral
$2048\times2048$ pixel Charge-Coupled Device (CCD) giving a pixel scale of
0.188\arcsec.  The ordinary and extraordinary beams appear as images
separated by about 15\arcsec\ on the CCD detector.

Observations began 21.88 hours 
after the burst with observing and reduction procedures similar 
to that used for GRB~990123 \citep{hjorth99}, each 
image providing fluxes in two of the four orientations required. 
Individual exposure times were 600 or 900 s resulting in 
total exposure times of 2400 s per orientation at a mean epoch of 
Feb 23.25169 2001 UT, 22.65 hours after the burst (see Table 1). 
No information on the position angle is available, as 
no polarization standards were observed. 
The images were flat fielded using standard procedures with appropriate
dome flats in each of the two orientations. The three images in each
orientation were combined. Absolute photometric calibration was performed
relative to star A of Henden et al.\ (2002) which has
$V=14.582\pm 0.022$. Meteorological conditions were fine and the 
measured resolution of the images was $\sim$ 0.8--$1.0\arcsec$. 
Figure~1a shows an excerpt of the combined images.

Aperture photometry was performed on the combined images with the
DAOPHOT~II/ALLSTAR (Stetson 1987, 1994) software package.
Point-spread function (PSF) photometry turned out to be difficult due to
the lack of suitable PSF stars and the strongly orientation-dependent PSF
shape and width. Aperture growth curves were computed for the OA,
star B and a galaxy (G) in the field (see Fig.~1a). Star A was
saturated in these images. Aperture photometry of star A was therefore
performed on the short exposures obtained immediately after the
six long exposures (cf.~Table~1).

In order to correct for possible interstellar polarization, polarization
induced by the telescope and instrument, and the significant variations
in stellar shape as a function of orientation, we computed aperture growth
curves for OA, B and G relative to A.

The Stokes vector, the instrumental Q and U values, and the linear
polarization were computed from the derived fluxes in the four orientations\\
\citep{hjorth99, wijers99, rol00} as a function of aperture radius.
The results are dependent on the aperture radii used. 
For small radii the results are dominated by photon noise and small 
variations in the shapes
and widths of the objects relative to star A.  For large radii the noise
from and difficulty in subtracting the sky background dominates. Thus,
one can only hope for reliable results for intermediate aperture radii.
We used two criteria for selecting the aperture radius, namely
(i) minimizing photon noise which favored aperture radii between 4 and 6
pixels and (ii) insensitivity to the aperture radii of the Q and U values
which favored aperture radii between 5 and 7 pixels. We note that B never
reached independence of aperture radius and shows a variation in Q
larger than that expected from photon noise. We therefore consider the
results for B unreliable, probably due to relatively high photon noise
from the fainter star. Aperture radii between
5 and 7 pixels for the OA and G are shown as filled circles in Fig.~1b. 
The mean of these give $P(OA) = 1.50\pm0.64\%$ and
$P(G) = 0.4 \pm 0.9\%$. We note that the polarization of the galaxy is
consistent with zero as it should if star A is unpolarized. 

The effects of depolarization or interstellar polarization in the 
Milky Way are expected to be negligible at the high latitude of 
GRB~010222 ($b=60.9^\circ$) --
well out of the Galactic plane. For the low Galactic extinction
of E(B$-$V) = 0.023 (Schlegel, Finkbeiner \& Davis 1998),
the maximum interstellar polarization or depolarization
is less than 0.2\% (Serkowski, Mathewson \& Ford 1975; Berdyugin \& Teerikorpi 1997). 
Thus the interstellar polarization or depolarization 
towards GRB 010222 is negligible.

At low significance levels of the computed afterglow polarization,
a correction must be applied to account for the non-Gaussian nature of
the underlying probability distribution \citep{wardle74}.
When corrected for this effect the $V$-band linear polarization of the optical
afterglow of GRB~010222 is $1.36\pm0.64$\%, similar to the degree of
polarization observed in a number of other bursts \citep{covino99, rol00}.

As seen in Table 1, the optical afterglow decayed during the observations.
As a check on the reliability of our polarization measurements we computed 
the V-band magnitudes by taking the mean of the two orientations at each 
epoch. The resulting decay slope of $-1.19\pm0.25$ is consistent with 
the observed lightcurve (e.g.\ Henden et al.\ 2002).

\subsection{X-rays}

We have also analyzed data from {\sl Chandra X-ray Observatory} 
observations starting about 15 hours after the burst, yielding a net 
exposure time of 29.5~ksec. Due to the brightness of the $X$-ray 
afterglow the data are severely affected by pile-up. Hence, the spectra 
were analyzed following the procedure suggested by Davis \cite{davis01}, 
taking pile-up effects explicitly into account in the spectral modeling. 
We extracted a spectrum from a 
circular 4 arcsec diameter region centered on the $X$-ray afterglow and 
fitted the 0.5-10 keV spectrum with three model components: i) An intrinsic 
power law, ii) Galactic absorption fixed at the nominal value 
$n_H=1.6 \times 10^{20}$~cm$^{-2}$ \citep{dl90}, and iii) intrinsic absorption in the GRB 
host fixed at the anticipated redshift of $z=1.477$. This model provides 
an excellent fit ($\chi^2=116$ for 204 d.o.f.), with a best fit spectral 
index $\beta=-0.72\pm 0.17$ and intrinsic GRB host absorption 
$n_H^i=6.5\pm 0.11 \times 10^{21}$~cm$^{-2}$ ($1\sigma$ errors). 
There is no evidence for additional spectral features. The 4 keV flux
at the mean epoch of observations, $0.81^d$, was $0.268\pm0.054$ $\mu$Jy. 
The near contemporaneous BeppoSAX observations give $\beta=-0.97\pm 0.05$
\citep{intzand01}. The difference between our spectral index and the BeppoSAX
results is due to different derived values of $n_H$. Using their method and 
fixing $n_H$ at their value, we find $\beta = -1.13\pm 0.20$.

\section{A Jet with Continuous Energy Injection}
\label{sec:cont}

Sources of synchrotron radiation are generally expected to exhibit
polarization in their emission, with the degree of polarization as
high as $60-70\%$ (e.g.\ Rybicki \& Lightman 1985). This is, however, 
strongly dependent on the degree of regularity in the magnetic field.
In fireball models, the field is expected to be highly entangled 
with no preferred direction, hence little or no polarization is expected. 
This would also be the case in a spherical fireball, due to symmetry,
even if the magnetic field had a regular component.

In a jet-like fireball, if the magnetic field has a regular 
component and the line of sight is at an angle with the center of 
collimation, some polarization may be observed \citep{gl1999, sari99}.
The degree of polarization observed in GRB~010222 thus suggests that 
the fireball is slightly asymmetric. It follows that a transition to a 
non-relativistic regime of a spherical fireball or a wide jet, 
is not likely to be a correct description of this event.

On the other hand, the observed light curve break magnitude, 
$\Delta\alpha=0.74\pm 0.04$, is exactly what is expected in a jet of 
fixed opening angle \citep{mesz99}. We will henceforth adopt that geometry.

The isotropic energy release of GRB~010222, as estimated from the SAX 
data is $E_{52}=154.2\pm17.0$  (in units of $10^{52}$ erg), in a cosmology 
with $\Omega_{\rm m}=0.3$, $\Omega_\Lambda=0.7$ and $h_0=0.65$  
\citep{amati02}. The initial Lorentz-factor, 
estimated from equation (10) in Sari \& Piran \cite{sarpir99}, 
is then $\Gamma_0\approx 350$, ignoring the weak dependence on the ambient 
number density, $n_0$. Using the light curve break time $t_b=0.58\pm 0.04$ 
days \citep{henden02}, we find that the corresponding collimation angle at 
$t_b$ is $\theta_0=1/\Gamma_b\approx 2.0^\circ$, reducing the estimate of the energy 
released in the burst to $E\approx 10^{51}$ ergs, fully consistent with
the results of Frail et al.\ \cite{frail01}, although near the higher
end of their distribution. 

We advance the hypothesis that the exceptionally slow light curve decay 
of GRB~010222 is the consequence of continuous energy injection. If the 
fireball energy injection is continuous over an extended period of time 
rather than instantaneous, it may be modeled as a power law in ejected 
mass \citep{rees98, sari00}. 
The resulting time evolution of the Lorentz-factor can then be expressed 
as $\Gamma=\Gamma_0 (t/t_0)^{-(3-g)/(7+s)}$, where $s$ is the power law 
exponent of the ejected mass distribution and $g$ is the power law index 
of the ambient density distribution. We will henceforth assume a homogeneous 
environment ($g=0$). Then $s=1$ reproduces the instantaneous 
case as the energy release is then constant, while $s>1$ implies that 
the energy is dominated by material with low Lorentz-factors 
\citep{rees98}. We will take $s=2$ as a representative case of 
moderate energy injection \citep{sari00}. Knowing the hydrodynamic 
evolution allows one to derive the corresponding light curve and spectral 
evolution as in the standard fireball model with instantaneous energy
injection. General expressions, valid also for inhomogeneous external
medium, can be found in Sari \& M\'esz\'aros (2000). We will use their 
notation in what follows. It should be emphasized that a reverse shock 
is expected to produce substantial flux at low frequencies, especially 
at late times. 

Interpreting the optical and $X$-ray light curves as due to the forward
shock we find, using $\alpha_1=-0.66\pm 0.03$ and $s=2$, that 
\begin{equation}
p=\frac{3(s+1)-(7+s)\alpha_1}{6}=2.49\pm 0.05,
\end{equation}
in agreement with many other bursts. It follows that the intrinsic spectral 
index is $\beta=-(p-1)/2=-0.75\pm0.03$. This is fully consistent with the 
$X$-ray observations, if the cooling frequency, $\nu_c$ was above the 
$X$-rays at the time of observations, but it requires moderate extinction 
in the optical (Lee et al.\ 2001; Masetti et al.\ 2001). It provides a 
consistent picture of the broadband spectrum from optical through the 
$X$-rays. The uncorrected $V$-flux at 0.81$^{\rm d}$ is 45$\mu$J, using the 
temporal index $\alpha=-1.19$ as inferred from our polarimetry. 
Adopting the SMC extinction correction of Lee et al.\ (2001), we
find that $A_V=0.3$ corresponds to $\beta=-0.72$, consistent with
the intrinsic spectral slope we inferred above and our $X$-ray 
observations.

We note that if $2\leq p \leq 3$, when $s=2$, then the pre-break slope would 
be in the range $0.33 \leq -\alpha_1 \leq 1.0$. For a jet with fixed $\theta_0$, 
we would then have $1.08 \leq -\alpha_2 \leq 1.75$. It is not straight forward 
to estimate the post break slope in an expanding jet model, as it requires a 
knowledge of how the continuous energy injection modifies the jet evolution 
subsequent to the moment when $\Gamma < 1/\theta_0$. This is the subject of 
a separate paper (Bj\"ornsson et al.\, in preparation). We expect, however,
a tranistion period where the temporal slope depends both on $p$ and $s$.
The asymptotic decay slope will in the end approach $-p$, but this may not 
be reached in all cases before the optical transient fades below detectability
or the host galaxy starts to dominate the emission. The late time decay slope
derived by Fruchter et al.\ (2001), $\alpha_2=-1.7\pm 0.05$, may indeed indicate
that the afterglow continued to steepen at very late times. We suggest that the 
above scenario may also apply e.g.\ to GRB~020813 as its light curve decay 
is very similar to that of GRB~010222 (e.g.\ Bloom, Fox \& Hunt 2002; Gladders \&
Hall 2002).

The flux from the forward shock has a maximum at the frequency $\nu_m^f$,
that in this interpretation is below the optical. The spectral region from 
optical through $X$-rays is therefore represented by 
the same power-law, $F_\nu\propto\nu^{-0.75}$. From our $X$-ray data
at $0.81^{\rm d}$, we obtain
\begin{equation}
n_0^{1/2}\epsilon_{B,-2}^{7/8}\epsilon_{e,0.5}^{3/2}=2.53\cdot 10^{-5}.
\label{constraint1}
\end{equation}
To solve for the model parameters, we need two more constraints. The optical
and the $X$-rays are degenerate in the sense, that they are both between
$\nu_m^f$ and $\nu_c$. The reverse shock is expected to produce substantial 
emission around the frequency $\nu_m^r$, that is defined analogously to $\nu_m^f$,
and is related to it by $\nu_m^r=\nu_m^f/\Gamma^2$, the peak flux of the reverse 
shock being a factor $\Gamma$ larger than the peak flux of the forward shock.
For $\nu_m^f\approx 10^{14}$ Hz,
$\nu_m^r$ could be from a few hundred GHz to $10^{12}$ Hz. Furthermore the flux
from the reverse shock may be expected to stay constant or even increase
for a period of time (Panaitescu, M\'esz\'aros \& Rees 1998), or decay 
as $t^{-1/2}$ \citep{sari00}. The radio and sub-mm observations of Frail et al.\ 
\cite{frail02}, show a constant flux onwards from about $0.35^{\rm d}$ and 
$0.8^{\rm d}$ at 350 and 250 GHz, respectively. They convincingly argue that
this flux originates in the host galaxy rather than the afterglow. That sets 
an upper limit on the emission predicted by the model, in principle enabling 
us to further constrain the parameters. 
In practice, as $\nu_m^f$ is unknown, the constraints in this case are very 
weak. It is apparent though from eq.\ (\ref{constraint1}), that for reasonable
values of $n_0$, and $\epsilon_{e,0.5}$, the magnetic field is very weak.

A generic property of extended injection models is that the afterglow
fades more slowly than in the case of instantaneous injection. We have 
chosen a simple power-law distribution of $\Gamma$ as an example of
such a scenario. Other, perhaps more detailed models, are of course 
possible (e.g.\ Zhang \& M\'esz\'aros 2002), but would only 
affect the above interpretation in the details. The interpretation 
presented here, requires the introduction of an additional model 
parameter ($s$), but this is also true in other models for the
case of $p<2$ as discussed above. An inferred value of $p>2$ is 
furthermore supported by detailed modelling of particle acceleration 
in relativistic shocks \citep{achterberg01}.

\acknowledgements
This research was supported in part by a {\em Special Grant} from the 
Icelandic Research Council, the University of Iceland Research Fund and 
the Danish Natural Science Research Council (SNF). We thank Davide Lazzati 
for constructive comments on an early version of the manuscript. We acknowledge 
the First Niels Bohr Summer Institute for hospitality during the completion of 
this work. GB thanks Einar H.\ Gudmundsson for a number of helpful discussions.


\begin{table}
\begin{center}
\caption{Journal of polarimetry observations of GRB~010222}
\begin{tabular}{lccc}
Time            \dotfill& orientation  & exp.~time & $V$        \\
(Feb 2001 UT)\dotfill&  (deg)       & (s)       & (mag)      \\
\hline
23.22499            \dotfill&  $45/135$   & 900 & $20.423\pm0.011$ \\
23.23640            \dotfill&  $0/90$     & 900 & $20.444\pm0.011$ \\
23.24782            \dotfill&  $45/135$   & 900 & $20.436\pm0.011$ \\
23.25925            \dotfill&  $0/90$     & 900 & $20.468\pm0.011$ \\
23.26894            \dotfill&  $45/135$   & 600 & $20.484\pm0.014$ \\
23.27688            \dotfill&  $0/90$     & 600 & $20.499\pm0.016$ \\
23.28154            \dotfill&  $0/90$     & 30  &                  \\
23.28289            \dotfill&  $45/135$   & 30  &                  \\

\hline
\label{table1}
\end{tabular}
\end{center}
The photometric zero point was determined assuming
$V=14.582$ for star A (cf.\ Henden et al.\ 2002).
\end{table}


\begin{figure}[t]
\plottwo{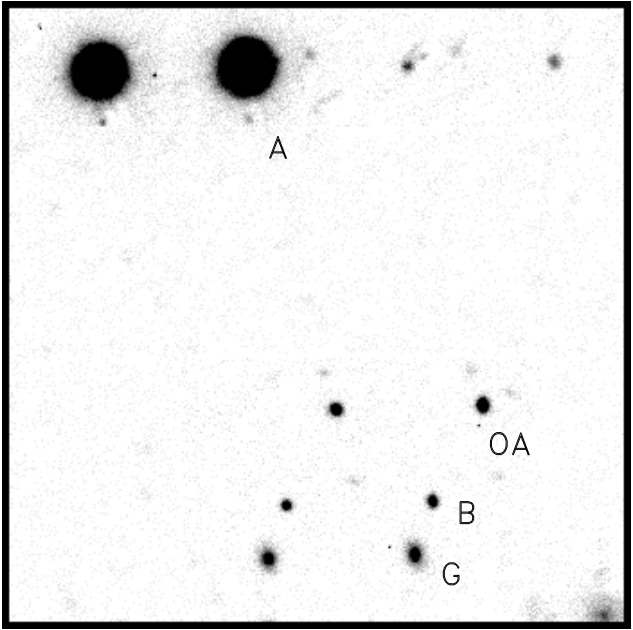}{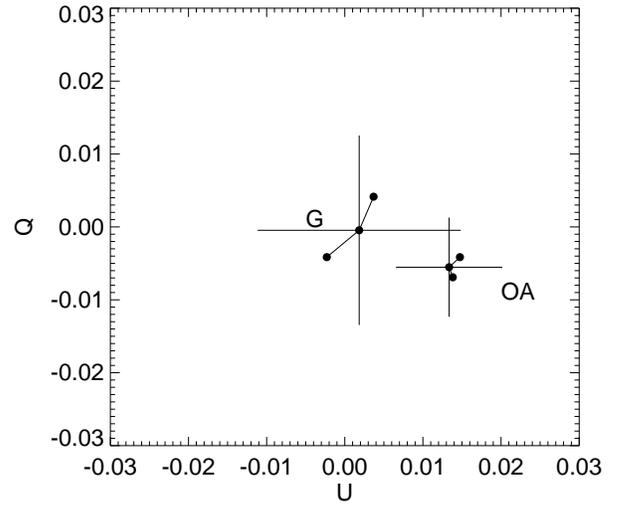}
\caption {a) An excerpt of the combined image
containing the $0^\circ/90^\circ$ orientations. The image measures $60\times 85$ 
arcsec. North is up and East is to the left. The optical afterglow of GRB~010222
(OA), Henden's bright star (A), a faint star (B) and a galaxy (G) are indicated. 
The orientations corresponding to $0^\circ$ and $90^\circ$ are separated by 
15 arcsec in the East-West direction.
b) This figure shows Q vs.\ U for the OA and G. Aperture radii 
of 5,6 and 7 pixels are plotted with photon noise error bars shown 
for 6 pixels. }
\end{figure}


\begin{thebibliography}{}

\bibitem[Achterbeg et al.\ 2001]{achterberg01} Achterbeg, A., Gallant, Y.\ A., 
Kirk, J.\ G., Guthmann, A.\ W., 2001, MNRAS 328, 393

\bibitem[Amati et al.\ 2002]{amati02} Amati, et al.\ 2002, A\&A 390, 81

\bibitem[Andersen et al.\ 1999]{andersen99} Andersen, M.\ I., et al. 
1999, Science 283, 2075

\bibitem[2001]{batta01} Bhattacharya, D., 2001, Bull.\ Astr.\ Soc.\ India, 29, 107

\bibitem[2002]{bloom02} Bloom, J.\ S., Fox, D.\ W.\ \& Hunt, M.\ P., 2002, GCN, 1476 

\bibitem[Berdyugin \& Teerikorpi 1997]{berdyugin97} Berdyugin, A., Teerikorpi, P.,
1997, A\&A 318, 37
 
\bibitem[Cowsik et al.\ 2001]{cowsik01} Cowsik, R., Prabhu, T. P., Anupama, G. C., 
Bhatt, B. C., Sahu, D. K., Ambika, S., Padmakar, Bhargavi, S. G.\
2001, Bull.\ Astr.\ Soc.\ India, 29, 157

\bibitem[e.g.\ Covino et al.\ 1999]{covino99} Covino, S., et~al.\ 1999, A\&A 348, L1

\bibitem[2001]{daicheng01} Dai, Z.\ G. \& Cheng, K.\ S.,
2001, ApJ 558, L109

\bibitem[2001]{davis01} Davis, J.\ E.,
2001, ApJ 562, 575

\bibitem[Dickey \& Lockman 1990]{dl90} Dickey, J.\ M.\ \& Lockman, F.\ J., 1990, ARA\&A, 28, 215

\bibitem[2001]{frail01} Frail, D.\ A., et al.\ 2001, ApJ 562, L55

\bibitem[2002]{frail02} Frail, D.\ A., et al.\ 2002, ApJ 565, 829

\bibitem[2001]{frucht01} Fruchter, A., Burud, I., Rhoads, J., Levan, A.\
2001, GCN, 1087

\bibitem[Ghisellini \& Lazzati 1999]{gl1999}
Ghisellini, G.\ \& Lazzati, D.\ 1999, MNRAS, 309, L7

\bibitem[2002]{gh02} Gladders, M.\ \& Hall, P., 2002, GCN, 1514

\bibitem[2001]{henden01} Henden, A.\ A.\ \& Vrba, F.\
2001, GCN, 967

\bibitem[Henden et al.\ 2002]{henden02} Henden, A.\ A., et al.,
2002, in preparation

\bibitem[Hjorth et al.\ 1999]{hjorth99} Hjorth, J., et al.\
1999, Science 283, 2073

\bibitem[2001]{jha01} Jha, S., et al.\ 2001, ApJ 554, L155

\bibitem[Kulkarni et al.\ 1999]{kulkarni99} Kulkarni, S.\ R., et al.,
1999, Nature 398, 389

\bibitem[2001]{lee01} Lee, B.\ C., et al.\
2001, ApJ 561, 183

\bibitem[2001]{masetti01} Masetti, N., et al.\ 2001, A\&A 374, 382

\bibitem[M\'esz\'aros 2002]{mesz02}
M\'esz\'aros, P., 2002, ARA\&A in press (astro-ph/0111170)

\bibitem[M\'esz\'aros \& Rees 1999]{mesz99} M\'esz\'aros, P., \& Rees, M.\ J., 
1999, MNRAS 306, L39

\bibitem[2002]{mirabal02} Mirabal, N., et al.\
2002, astro-ph/0207009

\bibitem[1998]{pmr98} Panaitescu, A.,  M\'esz\'aros, P., \& Rees, M.\ J., 
1998, ApJ 503, 314

\bibitem[2002]{pankum02} Panaitescu, A., \& Kumar, P., 
2002, ApJ, 571, 779

\bibitem[Piro et al.\ 2001]{piro01} Piro, L.\
2001, GCN, 959

\bibitem[2002]{price02} Price, P.\ A., et al.,
2002, astro-ph/0203467

\bibitem[Rees \& M\'esz\'aros 1998]{rees98} Rees, M.\ J., \& M\'esz\'aros, P.,
1998, ApJ 496, L1

\bibitem[Rol et~al.\ 2000]{rol00}Rol, E., et~al.\ 
2000, ApJ, 544, 707

\bibitem[1985]{ryblight85}Rybicki, G.\ B., Lightman, A.\ P., Radiative Processes
in Astrophysics, Wiley-Interscience, 1985

\bibitem[2001]{sagar01} Sagar, R., et al.\
2001, Bull.\ Astr.\ Soc.\ India. 29, 91

\bibitem[1998]{sari98a} Sari, R., Piran, T., \& Narayan, R.,
1998, ApJ 497, L17

\bibitem[1998]{sari98b} Sari, R., Piran, T., \& Halpern, J.\ P.,
1999, ApJ 519, L17

\bibitem[Sari 1999]{sari99}
Sari, R.\ 1999, ApJ, 524, L43

\bibitem[1999]{sarpir99} Sari, R., \& Piran, T.,
1999, ApJ 520, 641

\bibitem[Sari \& M\'esz\'aros 2000]{sari00} Sari, R., \& M\'esz\'aros, P.,
200, ApJ 535, L33

\bibitem[1998]{schlegel98} Schlegel, D.J., 
Finkbeiner, D.\ P., \& Davis, M.,
1998, ApJ 500, 525

\bibitem[1975]{serkowski75} Serkowski, K., Mathewson, D.\ S., \& Ford, V.\ L.,
1975, ApJ 196, 261

\bibitem[2001]{stanek01} Stanek, K.\ Z., et al.\ 2001, ApJ 563, 592

\bibitem[1987]{stetson87} Stetson, P.\ B.,
1987, PASP, 99, 191

\bibitem[1994]{stetson94} Stetson, P.\ B.,
1994, PASP, 106, 250

\bibitem[in 't Zand et al.\ 2001]{intzand01} in 't Zand, J.\ J.\ M., et al.\
2001, ApJ 559, 710

\bibitem[2002]{zm02} Zhang, B., \& M\'esz\'aros, P., 
2002, ApJ 566, 712

\bibitem[2000]{vp2002} van Paradijs, J., Kouvelioutu, C. \& Wijers, R.\ A.\ M.\ J.,
2000, ARA\&A, 38, 379

\bibitem[Wardle \& Kronberg 1974]{wardle74} Wardle, J.\ F.\ C., Kronberg, P.\ P., 1974, ApJ 194, 249

\bibitem[Wijers et al.\ 1999]{wijers99}
Wijers, R.\ A.\ M.\ J., et~al.\ 1999, ApJ, 523, L33

\end{thebibliography}
\end{document}